\documentclass[superscriptaddress,showkeys,showpacs,preprintnumbers,floatfix,twocolumn]{revtex4-1}
\pdfoutput=1
\usepackage{epsfig}
\usepackage{graphicx}

\usepackage{pslatex}
\usepackage{color}
\usepackage{amssymb}
\usepackage{multirow}
\usepackage{xspace}

\usepackage{float}

\usepackage{amssymb}
\usepackage{lineno}
\usepackage[figuresright]{rotating}
\newcommand{\Ups}{\ensuremath{\Upsilon}\xspace}
\newcommand{\UpsAll}{\ensuremath{{\Upsilon({1S+2S+3S})}}\xspace}
\newcommand{\UpsExc}{\ensuremath{{\Upsilon({2S+3S})}}\xspace}

\newcommand{\UpsOne}{\ensuremath{{\Upsilon({1S})}}\xspace}
\newcommand{\UpsTwo}{\ensuremath{{\Upsilon({2S})}}\xspace}
\newcommand{\UpsThree}{\ensuremath{{\Upsilon({3S})}}\xspace}
\newcommand{\Jpsi}{\ensuremath{J/\psi}\xspace}
\newcommand{\pT}{\ensuremath{p_\mathrm{T}}\xspace}

\newcommand{\GeV}{\ensuremath{\mathrm{GeV}}\xspace}
\newcommand{\gev}{\GeV}
\newcommand{\gevc}{\GeV/$c$}

\newcommand{\Raa}{\ensuremath{R_\mathrm{AA}}\xspace}
\newcommand{\Npart}{\ensuremath{N_\mathrm{part}}\xspace}
\newcommand{\Ncoll}{\ensuremath{N_\mathrm{coll}}\xspace}

\newcommand{\bb}{\ensuremath{b\bar{b}}\xspace}

\newcommand{\sqrtsNN}{\ensuremath{\sqrt{s_{\rm{NN}}}}}

\begin{document}
\title{$\Upsilon$ Production in Heavy-Ion Collisions from the STAR Experiment}


\author{Zaochen Ye for the STAR Collaboration}
\address{University of Illinois at Chicago, Chicago, Illinois, 60607, USA}
%%%%%%%%%%%%%%%%%%%%%%%%%%%%%%%%%%%%%%%%%%%%
\begin{abstract}
In these proceedings, we present recent results of $\Upsilon$ measurements in heavy-ion collisions from the STAR experiment at RHIC. Nuclear modification factors ($R_{AA}$) for $\Upsilon(1S)$ and $\Upsilon(1S+2S+3S)$ in U+U collisions at \sqrtsNN\ = 193 GeV are measured through the di-electron channel and compared to those in Au+Au collisions at \sqrtsNN\ = 200 GeV and Pb+Pb collisions at \sqrtsNN\ = 2.76 TeV. The ratio between the $\Upsilon(2S+3S)$ and $\Upsilon(1S)$ yields in Au+Au collisions at \sqrtsNN\ = 200 GeV is measured in the di-muon channel and compared to those in p+p collisions and in Pb+Pb collisions at \sqrtsNN\ = 2.76 TeV. Prospects for future $\Upsilon$ measurements with the STAR experiment are also discussed. 

\end{abstract}
%%%%%%%%%%%%%%%%%%%%%%%%%%%%%%%%%%%%%%%%%%%%

\keywords{Quark-Gluon Plasma (QGP), Color screening, Dissociation, Suppression, Upsilon,  STAR} 
\maketitle

%\linenumbers
\section{Introduction}
\label{introduction}
Quark-Gluon Plasma (QGP), a new state of matter where quarks and gluons are de-confined, is believed to have existed up to a few milliseconds after the Big Bang. Quarkonia could dissociate in the QGP due to color screening of quark-antiquark potential by the surrounding partons in the medium \cite{Matsui:1986dk}, which was suggested as a signature of QGP formation in heavy-ion collisions. Moreover, different quarkonium states may dissociate at different temperatures depending on their binding energies~\cite{Digal:2001iu, Wong:2004zr, Cabrera:2006nt}. This so-called ``sequential melting" phenomenon could be used to deduce the temperature of the QGP. However, other effects, such as regeneration from deconfined heavy quark-antiquark pairs, shadowing and antishadowing of nuclear parton distribution functions and co-mover absorption, need to be taken into account when interpreting experimental results. Compared to charmonium production at RHIC energies, bottomonium production has several advantages: 1) the regeneration contribution is negligible due to the the much smaller $b\bar{b}$ production cross section ($\sigma_{b\bar{b}} \approx 1.87_{-0.67}^{+0.99}$ $\mathrm{\mu b}$  \cite{Cacciari:2005rk}
compared to $\sigma_{c\bar{c}} \approx 550-1400$ $\mathrm{\mu b}$  \cite{Adams:2004fc} ); 2) the cross section for inelastic collisions of \Ups\ with hadrons is very small, hence the co-mover absorption is predicted to be minimal \cite{Lin:2000ke}; 3) the suppression of \Ups\ production due to cold-nuclear-matter (CNM) effects has been measured to be smaller than that for $J/\psi$ reported by NA50 \cite{Alessandro:2006jt}. Thus, the \Ups\ family is expected to be a cleaner and more direct probe of the QGP, and the corresponding color deconfinement effects. \\

$\Upsilon$ production has been studied via the di-electron decay channel at STAR in different collision systems, including  p+p, d+Au and Au+Au  collisions at \sqrtsNN\ = 200 GeV~\cite{Adamczyk:2013poh}. The latest \Ups\ measurement via the di-electron channel in U+U  collisions at \sqrtsNN\ = 193 GeV allows a study of \Ups\ suppession in a new heavy-ion collision system \cite{Adamczyk:2016dzv}. Since 2014, a new detector, the Muon Telescope Detector (MTD), has been fully installed and taking data, allowing measurements of \Ups\ production via the di-muon channel. Compared to the di-electron channel, the di-muon channel has better sensitivity to different Upsilon states due to the reduced bremsstrahlung radiation. 

\section{\Ups\ \Raa via the di-electron channel in U+U and Au+Au collisions }

$\Upsilon \rightarrow e^{+}e^{-}$ decays were reconstructed using the Time Projection Chamber (TPC) and Barrel ElectroMagnetic Calorimeter (BEMC) with full azimuthal coverage over the pseudorapidity range  $|\eta|<1$. Electron identification (eID) was achieved by measuring the ionization energy loss ($dE/dx$) and track momentum by the TPC, as well as the energy deposition in the BEMC. In addition, shower profiles measured by the Barrel Shower Maximum Detector (BSMD) were used in Au+Au collisions to further suppress hadron contamination. The identified electron and positron candidates are paired to reconstruct the invariant mass of the \Ups candidates. 

%%% invariant masses
\begin{figure}[H]
  \includegraphics[scale=0.45]{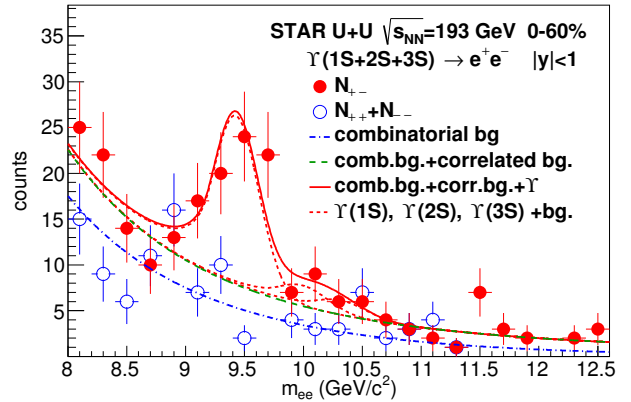}
  \caption{\label{fig:invmass} {\it (Color online)}     
Reconstructed invariant mass distribution of \Ups{} candidates \cite{Adamczyk:2016dzv}, Fits to the combinatorial background, \bb and Drell-Yan contributions
and to the \Ups peaks are plotted as dash-dotted, dashed and solid lines respectively. The fitted contributions of the individual \UpsOne, \UpsTwo and \UpsThree states are shown as dotted
lines.}
\end{figure}

\begin{figure}[th]
 \includegraphics[scale=0.45]{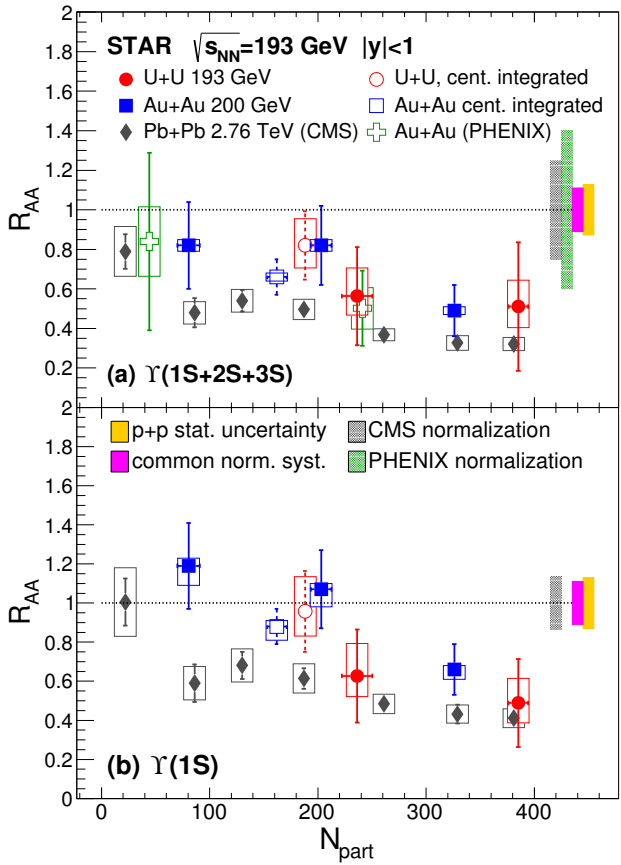}
  \caption{\label{fig:raa-data} {\it (Color online)}
   \UpsAll (a) and \UpsOne (b) \Raa vs \Npart in \sqrtsNN\ =193 GeV U+U collisions (solid circles) \cite{Adamczyk:2016dzv}, compared to 200 GeV RHIC Au+Au (solid squares~\cite{Adamczyk:2013poh} and hollow crosses~\cite{Adare:2014hje}), and 2.76 TeV LHC $\rm{Pb}+\rm{Pb}$ data (solid diamonds~\cite{Chatrchyan:2012lxa}). Each point is plotted at the center of its bin. Centrality integrated (0-60\%) U+U and Au+Au data are also shown as open circles and squares, respectively.}
\end{figure}

The \UpsAll and \UpsOne $R_{AA}$ in U+U collisions at \sqrtsNN\ = 193 GeV were calculated by dividing the invariant $\Upsilon$ yields in U+U collisions by those in p+p collisions scaled by the number of binary nucleon-nucleon collsions (\Ncoll)  \cite{Adamczyk:2016dzv}. They are shown as a function of the number of participating nucleons (\Npart) in Fig.~\ref{fig:raa-data} and compared to those in Au+Au collisions at \sqrtsNN\ = 200 GeV within $|y|<1$ from STAR~\cite{Adamczyk:2013poh} , within $|y|<0.35$ from PHENIX~\cite{Adare:2014hje}, and in $\rm{Pb}+\rm{Pb}$ collisions at \sqrtsNN\ = 2.76 TeV within $|y|<2.4$ from CMS~\cite{Chatrchyan:2012lxa}. \UpsAll suppression becomes significant only in the most central collisions at RHIC energies. 
After combining U+U and Au+Au results, we find that $\Raa^\UpsOne=0.63 \pm 0.16 \pm 0.09$, which suggests that \UpsOne{} is significantly but not completely suppressed in central heavy-ion collisions at top RHIC energies. While both the RHIC and LHC data show suppression in the most central bins, $\Raa^\UpsOne$ is slightly, although not significantly, higher in semi-central collisions at RHIC than that at the LHC.

%%% Upsilon RAA vs Npart, data
%%% Upsilon RAA vs Npart, model
\begin{figure}[th]
\includegraphics[scale=0.45]{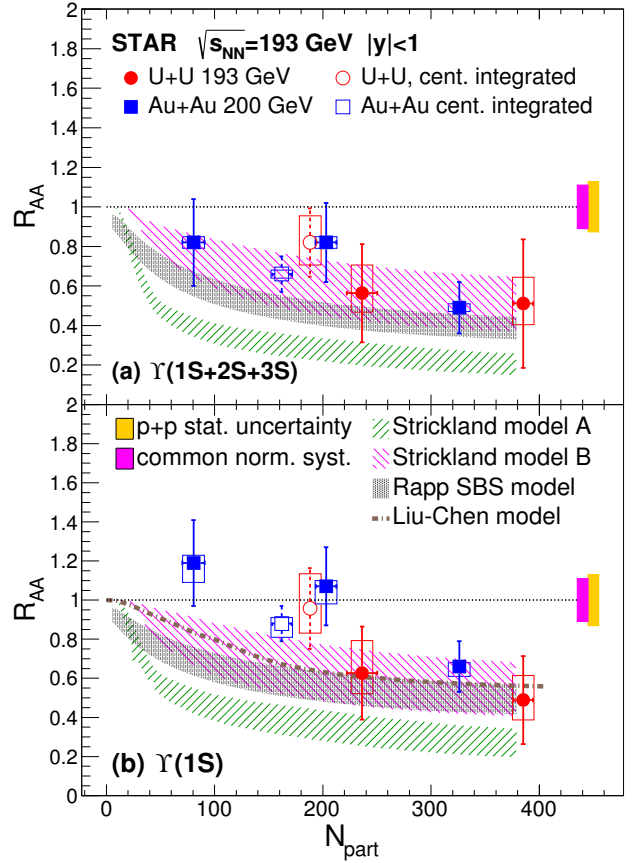}
  \caption{\label{fig:raa-model} {\it (Color online)}
   \UpsAll (a) and \UpsOne (b) \Raa vs. \Npart in $\sqrt{s_{NN}}=193$ GeV U+U collisions
   (solid circles) \cite{Adamczyk:2016dzv} and 200 GeV Au+Au collisions (solid squares), compared to different models~\cite{Emerick:2011xu,Strickland:2011aa,Liu:2010ej} described in the text. Each point is plotted at the center of its bin. Centrality integrated (0-60\%) U+U and Au+Au data are also shown as open circles and squares, respectively.}
\end{figure}
%%% Quarkonium RAA vs Ebinding
\begin{figure}[th]
 \includegraphics[scale=0.40]{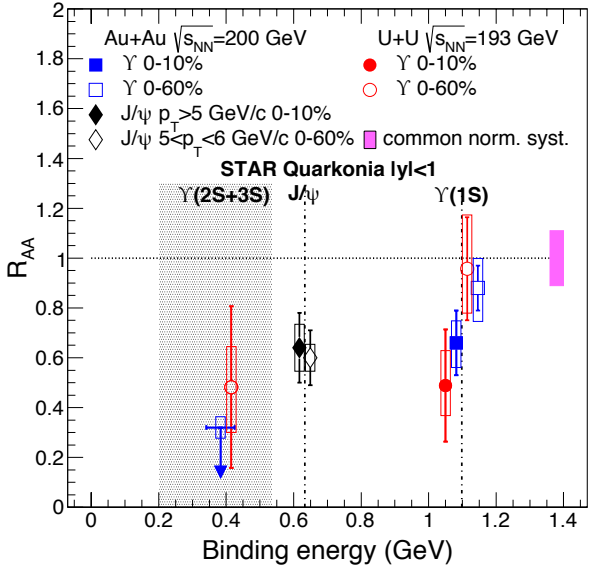}
  \caption{\label{fig:binding} {\it (Color online)}
   Quarkonium \Raa versus binding energy in Au+Au and U+U collisions. Open symbols represent 0-60\% centrality data, while filled symbols are for 0-10\% centrality. The \Ups measurements in U+U collisions are denoted by red points. In the case of Au+Au collisions, the \UpsOne measurement is denoted by a blue square, while for the \UpsExc states, a blue horizontal line indicates a 95\% upper confidence boundary. The black diamonds mark the high-\pT{} \Jpsi measurement. The data points are slightly shifted to the left and right from the nominal binding energy values to improve their visibility.
 } 
\end{figure}

In Fig.~\ref{fig:raa-model},  we compare STAR measurements to different theoretical models~\cite{Emerick:2011xu,Strickland:2011aa,Liu:2010ej}. An important source of uncertainty in model calculations for quarkonium dissociation stems from the unknown nature of the in-medium potential between the quark-antiquark pairs. Two limiting cases that are often used are the internal-energy-based heavy quark potential corresponding to a strongly bound scenario (SBS), and the free-energy-based potential corresponding to a more weakly bound scenario (WBS)~\cite{Grandchamp:2005yw}. The model of Emerick, Zhao and Rapp~\cite{Emerick:2011xu} includes CNM effects, dissociation of bottomonia in the hot medium (assuming a temperature of $T=330$ MeV) and regeneration for both the SBS and WBS scenarios. The Strickland-Bazow model~\cite{Strickland:2011aa} calculates dissociation in the medium in both a free-energy-based ``model A'' and an internal-energy-based ``model B'', with an initial central temperature
$428<T<442$ MeV. The model of Liu {\it et al.}~\cite{Liu:2010ej} uses an internal-energy-based potential and an input temperature $T=340$ MeV. In Fig.~\ref{fig:raa-model} we show all three internal-energy-based models together with the ``model A'' of Ref.~\cite{Strickland:2011aa} as an example for the free-energy-based models. The comparision between data and theoretical predictions suggests that internal-energy-based models generally describe RHIC data better than the free-energy-based models for the \UpsOne.

Figure~\ref{fig:binding} shows the \Raa versus binding energy of \UpsOne and \UpsExc states~\cite{Satz:2006kba} in U+U and Au+Au collisions. The results are also compared to high-\pT $J/\psi$ in Au+Au collisions~\cite{Adamczyk:2012ey}. This comparison is motivated by the expectation from model calculations, e.g. Ref.~\cite{Liu:2009nb}, that charm recombination is moderate at higher momenta. The results in U+U collisions are consistent with the Au+Au measurements as well as with the expectations from the sequential melting hypothesis. In the Au+Au data, the \UpsExc excited states have been found to be strongly suppressed, and a 95\% confidence upper limit $\Raa^\UpsExc < 0.32$ was established \cite{Adamczyk:2013poh}. The \UpsExc suppression observed in U+U data is consistent with this upper limit.

%%%%%%%%%%%%%%%%%%%%%%%%%%%%%%%%%%%%%%%%%%%%%%%%%%%
\section{$\Upsilon(2S+3S)/\Upsilon(1S)$ via the di-muon channel in Au+Au collisions}

The new STAR detector MTD was fully installed in 2014, allowing the \Ups reconstruction via the di-muon channel for the first time at STAR.  Muon candidates are identified using the TPC and MTD. Charged tracks are required to have \pT above 1.5 \gevc, and the differences between the measured and expected energy losses for muons are within $[-1\sigma, +3\sigma]$ where $\sigma$ is the $dE/dx$ resolution of the TPC. Tracks also need to geometrically match to the hits measured by the MTD, which covers about 45\% in azimuth within $|\eta|<0.5$. Cuts are applied to the residuals between projected track positions and MTD hit positions along both $z$ and $\varphi$ directions. In addition, the differences between the measured and expected time-of-flight for tracks from primary vertices to the MTD do not exceed 0.46 ns for accepted muon candidates. The identified muon candidates are then paired to reconstruct the invariant mass of the \Ups\ candiates. 

%%% Upsilon signal in di-muon channel
\begin{figure}[H]
  \includegraphics[scale=0.35, angle=270]{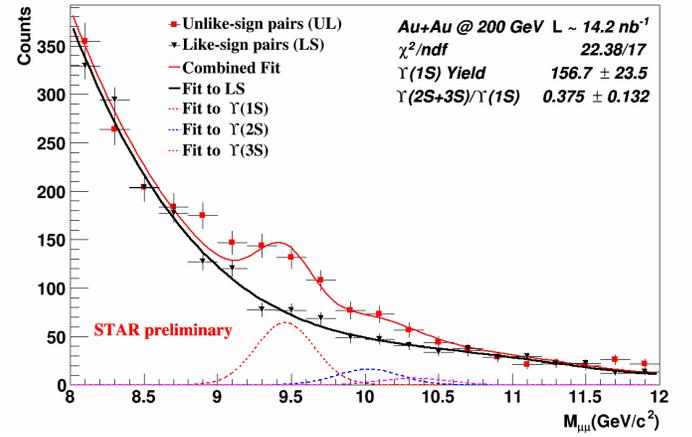}
  \caption{\label{fig:upsSig_mtd} {\it (Color online)}
              Invariant mass distribution of \Ups $\rightarrow$ $\mu^{+}\mu^{-}$ candidates for centrality 0-80\% in Au+Au collisions at 200 \gev. 
   }
\end{figure}

\begin{figure}[H]
  \includegraphics[scale=0.44]{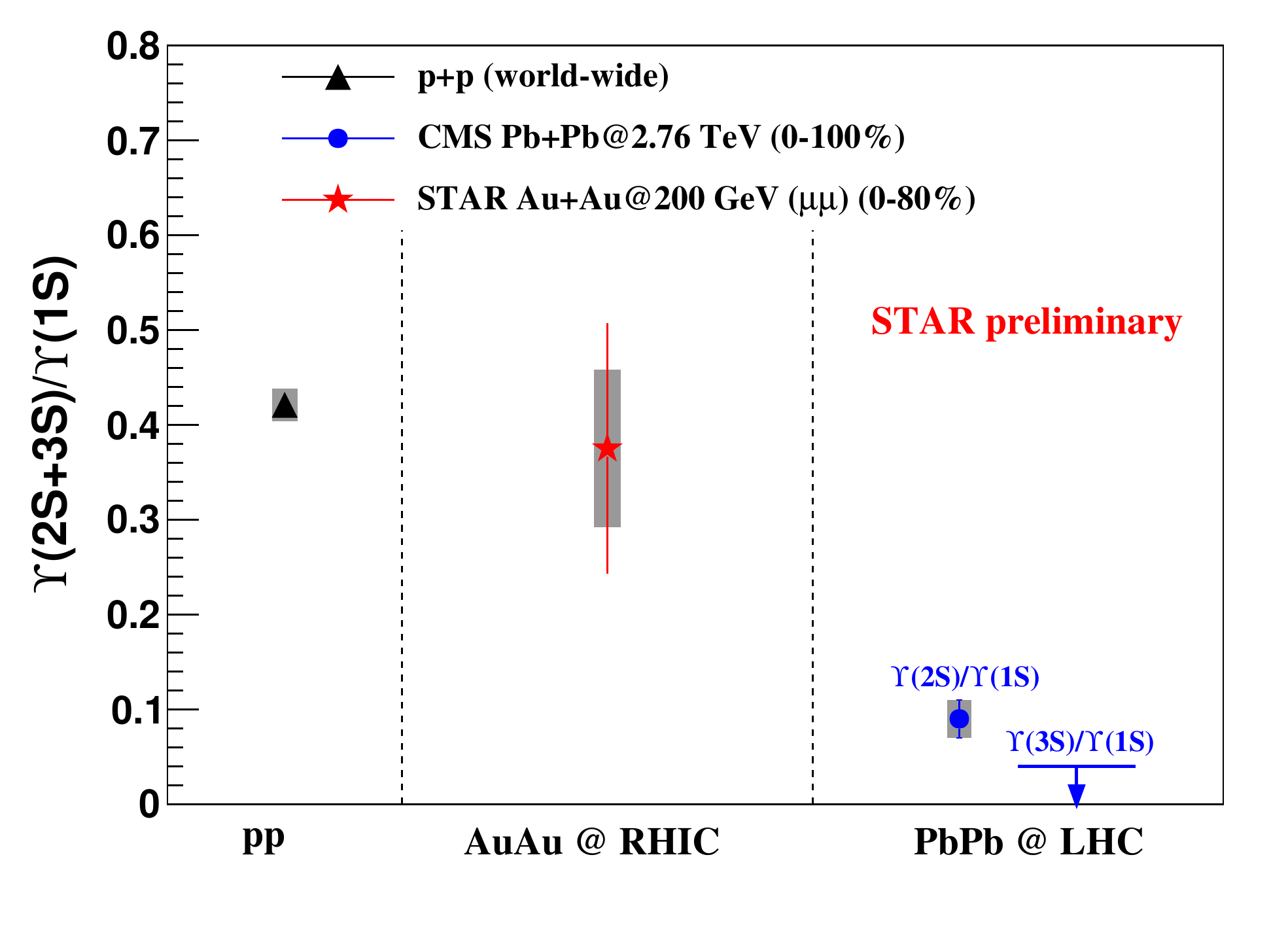}
  \caption{\label{fig:upsRatio_mtd} {\it (Color online)}
               \UpsExc/\UpsOne ratio via the di-muon channel, compared with world-wide measurement in p+p collisions \cite{Zha:2013uoa} and CMS measurements in $\rm{Pb}+\rm{Pb}$ collisions at \sqrtsNN\ = 2.76 TeV \cite{Chatrchyan:2012lxa,Chatrchyan:2013nza}.
  }
\end{figure}

 Figure~\ref{fig:upsSig_mtd} shows the di-muon mass spectrum in Au+Au collisions at \sqrtsNN\ = 200 GeV. The raw yields of \Ups\ states are obtained by a simultaneous fit to the like-sign and unlike-sign distributions. The \Ups\ state masses are fixed to the PDG values and their widths are determined by simulation. The ratio of \UpsTwo/\UpsThree is fixed to the value from world-wide meansurements in p+p collisions, and the shape of $b\bar{b}$ and Drell-Yan background is determined from PYTHIA.

Figure~ \ref{fig:upsRatio_mtd} shows the fitted \UpsExc\//\UpsOne\ ratio compared with the world-wide p+p data\cite{Zha:2013uoa}
and CMS data \cite{Chatrchyan:2012lxa,Chatrchyan:2013nza}. There is a hint of less melting of excited \UpsExc\ states relative to the ground \UpsOne\ state at RHIC than that at LHC. 

%%%%%%%%%%%%%%%%%%%%%%%%%%%%%%%%%%%%%%%%
%%%%%%%%%%%%%%%%%%%%%%%%%%%%%%%%%%%%%%%%
\section{Summary and Outlook}
%\begin{figure}
 % \includegraphics[width=\linewidth]{dAu.png}
  %\caption{\label{fig:dAu} {\it (Color online)}
     %          $R_{dAu}$ vs. $y$ for STAR (red stars) and PHENIX (green diamonds) results. The band on the right shows the overall normalization uncertainty for the STAR results due to systematic uncertainties in the p+p measurement. The shaded band shows the prediction for $R_{dAu}$ from EPS09 and its uncertainty. The dashed curve shows suppression due to initial-state parton energy  loss and the dot-dashed curve shows the same model with EPS09 incorporated \cite{Arleo:2012rs}.
  %}
%\end{figure}
\Ups\ production has been studied in different collision systems at the STAR experiment. A significant suppression of \Ups\ production at RHIC energies was observed in Au+Au and confirmed in U+U collisions at \sqrtsNN\ = 200 GeV and 193 GeV, respectively. Measurement of \Ups\ production via the di-muon channel indicates that the excited \UpsExc\ states are not completely suppressed in Au+Au collisions, and hints that the dissociation of \UpsExc\ relative to \UpsOne\ at RHIC energies is less than that in Pb+Pb collisions at \sqrtsNN\ = 2.76 TeV at the LHC. The new data taken in 2016 will double the data size for \Ups\ analysis in the di-muon channel, which may improve the precision of  \Ups\ measurement. The on-going analysis of \Ups\ measurements via the di-electron channel with futher optimized track quality cuts and the large data sets taken in 2011 and 2014 Au+Au runs may also allow the extraction of excited \Ups\ states. There are also on-going analyses with large-statistics data samples of p+p and p+Au collisions at \sqrtsNN\ = 200 GeV taken in 2015 which will greatly improve the \Ups\ reference measurements and provide more precise measurements of the CNM effects on the \Ups\ production at RHIC. 

%%%%%%%%%%%%%%%%%%%%%%%%%%%%%%%%%%%%%%%%%%%%%%%%%%%%%%%%
%%%%%%%%%%%%%%%%%%%%%%%%%%%%%%%%%%%%%%%%%%%%%%%%%%%%%%%%

\nocite{*}
\bibliographystyle{elsarticle-num}
\bibliography{jos}

\end{document}